\documentclass[aps,prl,twocolumn,showpacs,preprintnumbers,amsmath,amssymb]{revtex4-1}
 
\usepackage{latexsym,braket}

 \def\frac#1#2{{#1\over #2}}

\def\be{\begin{equation}}
\def\ee{\end{equation}}
\newcommand{\bea}{\begin{eqnarray}}
\newcommand{\eea}{\end{eqnarray}}

\newcommand{\SL}{\mathrm{SL}}
\newcommand{\mt}[1]{\textrm{\tiny #1}}

\usepackage{color}
\usepackage{graphicx}
\usepackage{dcolumn}
\usepackage{bm}
\usepackage{hyperref}

\begin{document}

\begin{flushright}
YITP-24-137
\end{flushright}
\title{Spread Complexity Rate as Proper  Momentum}

\author{Pawel Caputa$^{1,2,3}$, Bowen Chen$^{2,4}$, Ross W. McDonald$^{5}$, Joan Sim\'on$^{5}$, Benjamin Strittmatter$^{5}$}

\affiliation{
$^1$The Oscar Klein Centre and Department of Physics, Stockholm University, AlbaNova, 106 91 Stockholm, Sweden}
\affiliation{$^2$Faculty of Physics, University of Warsaw, Pasteura 5, 02-093 Warsaw, Poland }
\affiliation{$^3$Yukawa Institute for Theoretical Physics, Kyoto University, Kitashirakawa Oiwakecho, Sakyo-ku, Kyoto 606-8502, Japan}
\affiliation{$^4$Kavli Institute for Theoretical Sciences, University of Chinese Academy of Sciences, Beijing 100190, China}
\affiliation{$^5$School of Mathematics and Maxwell Institute for Mathematical Sciences, University of Edinburgh, Edinburgh EH9 3FD, UK}

\date{\today}

\begin{abstract}
We demonstrate a precise relation between the rate of complexity of quantum states excited by local operators in two-dimensional conformal field theories and the radial momentum of particles in 3-dimensional Anti-de Sitter spacetimes. Similar relations have been anticipated based on qualitative models for operator growth. Here, we make this correspondence sharp with two key ingredients: the precise definition of quantum complexity given by the spread complexity of states, and the match of its growth rate to the bulk momentum measured in the proper radial distance coordinate. 
\end{abstract}

\maketitle

\noindent \textbf{\emph{1.~Introduction.}} 
The concept of complexity, primarily used to quantify the difficulty of achieving tasks in information theory and computation \cite{shannon1949synthesis,kolmogorov1965three,rissanen1986stochastic,Nielsen:2005mkt,Ethan:2006gnj}, is gradually taking a central stage in physics. However, we expect that a physical definition of complexity should be associated with a number that could potentially be measured. Finding such definitions has been a focus of interdisciplinary research for many years  \cite{zurek2018complexity,Aaronson:2016vto,aharonov2011complexity,Osborne_2012,Caputa:2017yrh,Jefferson:2017sdb,Chapman:2017rqy,Caputa:2018kdj,Chagnet:2021uvi,Chen:2020nlj,Balasubramanian:2019wgd,Cotler:2017jue,Brandao:2019sgy,Haferkamp:2021uxo,Munson:2024usy,Brown:2021uov,Brown:2021euk,Bouland:2019pvu}.     

Surprisingly, complexity may also be the key to understanding black hole interiors. As first noticed in \cite{Susskind:2014rva}, the linear growth with time of generic black hole interiors, probed by extremal surfaces, might be a manifestation of their complexity and ability to process information. Such deep statements can only be tested within a quantum theory of gravity and, indeed, the AdS/CFT correspondence \cite{Maldacena:1997re} has already provided some supporting evidence \cite{Stanford:2014jda,Brown:2015bva,Belin:2021bga,Iliesiu:2021ari}. 

A concrete outcome of these developments is a conjecture stating that the rate of growth of complexity is proportional to the radial momentum of massive particles in Anti-de Sitter (AdS) spacetimes \cite{Susskind:2014jwa,Susskind:2018tei,Susskind:2019ddc,Brown:2018kvn,Susskind:2020gnl,Magan:2018nmu,Barbon:2020uux,Barbon:2020olv,Barbon:2019tuq,Lin:2019kpf,Lin:2019qwu,Ageev:2018msv,Ageev:2018nye} 
\begin{equation}
  \dot{C}(t) \equiv\frac{d}{dt}C(t) \propto P,
\label{eq:goal}  
\end{equation}
where $C(t)$ stands for a measure of complexity capturing the effective size of  Heisenberg operators $\mathcal{O}(t)\equiv e^{iHt}\,\mathcal{O}\,e^{-itH}$, whose dual description is given by falling particles with momentum $P$ in AdS. The main challenge in advancing this program has been the lack of a precise definition of quantum complexity (the left side of \eqref{eq:goal}) in quantum field theory (QFT), and matching it with momentum on the gravity side (the right side of \eqref{eq:goal}).

Since Heisenberg evolution of quantum operators is ubiquitous in physics, many new approaches to quantum complexity and operator growth have been developed \cite{Gopalakrishnan:2018wyg,Khemani:2018sdn,vonKeyserlingk:2017dyr,Parker:2018yvk}. When working on a 1D chain, there is a natural notion of size related to the support of a given operator as time flows. This picture was developed within the Sachdev-Ye-Kitaev (SYK) model \cite{Sachdev:1992fk,kitaev} in \cite{Qi:2018bje,Roberts:2018mnp} and relation \eqref{eq:goal} was confirmed within a gauge invariant formulation of a quantum point particle in AdS$_2$ \cite{Lin:2019qwu}. Still, it was not clear whether the SYK definition of size could be generalized to the higher-dimensional, interacting conformal field theories (CFTs) in conventional holography.

Important progress on defining operator complexity was achieved in many-body physics \cite{Parker:2018yvk} by using the Krylov basis and Krylov complexity (see e.g. \cite{Nandy:2024htc,Barbon:2019wsy,Rabinovici:2020ryf,Rabinovici:2022beu,Rabinovici:2022beu,Dymarsky:2021bjq,Jian:2020qpp,Hornedal:2022pkc,Basu:2024tgg,Balasubramanian:2022dnj,Balasubramanian:2023kwd,Sanchez-Garrido:2024pcy}). Furthermore, its dependence on the choice of inner product in the space of operators \cite{Magan:2020iac} was circumvented by introducing the spread complexity (reviewed below) of the unitary evolution of quantum states \cite{Balasubramanian:2022tpr}. The first tests of this new complexity measure in SYK models \cite{Lin:2022rbf,Rabinovici:2023yex} provided evidence of its relevance in holography and quantum gravity. 

In this letter, we report another milestone in this program by explicitly demonstrating \eqref{eq:goal} in an AdS/CFT set-up. Namely, we compute the rate of growth of the spread complexity for locally excited states in 2D CFTs and match it with the proper radial momentum of the dual massive particle in AdS$_3$. This elevates \eqref{eq:goal} from a low-dimensional observation to a new entry in the quantum information chapter of the holographic dictionary (see formula \eqref{eq:final}).\\
\vspace{4 pt}
\noindent \textbf{\emph{2.~Complexity of Spread of States.}}
The concepts of Krylov and spread complexity \cite{Balasubramanian:2022tpr} extend the intuitive notion of operator size on a lattice to arbitrary quantum systems. The key idea (called the recursion method, \cite{LanczosBook}) is to map the quantum dynamics to motion on an effective one-dimensional chain spanned by the Krylov basis, i.e. the subspace of orthonormal states $\{\ket{K_n}\}$ visited during the Hamiltonian evolution of the initial state
\begin{equation}
    \ket{\psi(t)}=e^{-iHt}\ket{\psi_0}=\sum_n \psi_n(t)\ket{K_n}\,.
\end{equation}
The Krylov basis is obtained by applying the Lanczos algorithm \cite{Lanczos:1950zz} to the Krylov subspace $\{\ket{\psi_0},H\ket{\psi_0},H^2\ket{\psi_0},...\}$. By construction, the coefficients in the Krylov basis, $\psi_n(t)$ satisfy a discrete Schr\"odinger equation
\begin{equation}
i\partial_t\psi_n(t)=a_n\psi_n(t)+b_n\psi_{n-1}(t)+b_{n+1}\psi_{n+1}(t), \label{SE1}
\end{equation}
controlled by the Lanczos coefficients $a_n$ and $b_n$. These are a byproduct of the Lanczos algorithm, and can also be extracted from the moments of the return amplitude  
\be
S(t)=\langle \psi(t)|\psi_0\rangle.
\label{eq:RetAmp}
\ee
The sites of the one-dimensional chain emerging in \eqref{SE1} are labelled by the Krylov index $n$, and the Lanczos coefficients describe the probability amplitude of staying on a given site ($a_n$) or hopping to the neighbouring sites ($b_n$). Solving \eqref{SE1} yields the probability distribution $p_n(t)=|\psi_n(t)|^2$, capturing the entire quantum dynamics. 

The spread complexity is defined as the average position on the 1D Krylov chain
\begin{equation}
    C_{\mt{K}}(t)=\sum_n n|\psi_n(t)|^2=\langle \psi(t)|\mathcal{K}|\psi(t)\rangle.\label{SpreadComplexity}
\end{equation}
where $\mathcal{K}=\sum_nn\ket{K_n}\bra{K_n}$ defines the ``complexity operator". This quantity is basis-dependent, in a similar way to some quantum coherence monotones \cite{PhysRevLett.113.140401}.
Eq. \eqref{SpreadComplexity} will be our definition of quantum state complexity. To compare it with momentum, we need a precise set-up that we introduce next.\\
\vspace{4 pt}
\noindent \textbf{\emph{3.~Set-up.}} 
The AdS/CFT correspondence states that the Hilbert space $\mathcal{H}_{\mt{CFT}}$ and the dynamics of certain (holographic) CFTs are equivalent to the Hilbert space $\mathcal{H}_{\mt{AdS}}$ and dynamics of excitations in AdS. In particular, for semi-classical states, such as the vacuum or thermal states, if one considers excitations generated by primary operators $\mathcal{O}$, their Heisenberg evolution $\mathcal{O}(t) = e^{iHt}\,\mathcal{O}\,e^{-itH}$ should be encoded in a relevant Krylov subspace of the full $\mathcal{H}_{\mt{CFT}}$. In a certain regime, one expects the gravitational dual description of such excitations to be approximately given by propagating bulk particles in an AdS geometry. The question we are asking is: \emph{how is the spread complexity of the operator in the CFT encoded in the motion of such a bulk particle?}

To make this quantitative, we consider the evolution of locally excited states by primary operators in holographic CFTs  \cite{Nozaki:2014hna,Caputa:2014vaa}
\begin{equation}
\ket{\psi(t)}=\mathcal{N}e^{-iHt}e^{-\epsilon H}\mathcal{O}(x_0)\ket{\psi_\beta},\label{LOES}
\end{equation}
where $\mathcal{N}$ is a normalisation of the state, $\mathcal{O}$ is a local primary operator of dimension $\Delta=h+\bar{h}$ inserted at spatial position $x=x_0$, $\beta=1/T$ is the inverse temperature, and $\ket{\psi_\beta}$ can be thought of as a purification of the thermal state (the thermofield-double state) \footnote{Results for the vacuum state in a CFT on the infinite line and on a circle of size $L$ are obtained by taking $\beta\to\infty$ and, formally, $\beta\to iL$, respectively. This is why we focus on finite $\beta$ in the main text. See supplemental material for further details.}.

The $\epsilon$-regularisation (smearing) in \eqref{LOES} is necessary for $\ket{\psi(t)} \in\mathcal{H}_{\mt{CFT}}$ and to have finite energy
\be
E_0=\int dx\langle \psi(t)| T_{00}(x) |\psi(t)\rangle =\frac{\Delta}{\epsilon}.
\label{eq:energy}
\ee
The CFT dynamics of this state describes an entangled pair of excitations propagating to the left and to the right of $x=x_0$ at the speed of light  \cite{Nozaki:2013wia}. This wave function spreading will be encoded by the Krylov probabilities $|\psi_n(t)|^2$ solving \eqref{SE1}. 
To study the validity of \eqref{eq:goal}, we need the bulk description of the state \eqref{LOES}. Next, we revisit its well-understood semi-classical approximation \cite{Nozaki:2013wia,Caputa:2014vaa,Asplund:2014coa}. 

\begin{figure}[t!]
    \centering
   \includegraphics[width=0.40\textwidth]{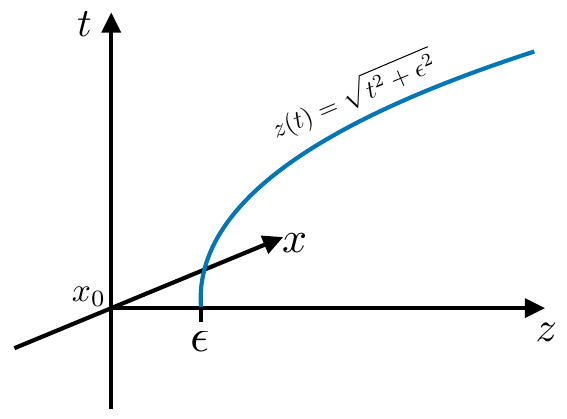}
    \caption{The holographic dual of a state excited by a local operator \eqref{LOES} corresponds to a massive particle following a timelike geodesic in $AdS_3$. Here we show the example of an excitation to the CFT vacuum on a line and the corresponding geodesic in \eqref{AdSPoincare}. }
    \label{fig:AdSP}
\end{figure}

Consider first the limit $\beta\to \infty$, so that $|\psi_\beta\rangle\to |0\rangle$ corresponds to the vacuum on the plane, whose semi-classical dual description is given by the Poincar\'e AdS$_3$ geometry
\be
ds^2=\frac{-dt^2+dx^2+dz^2}{z^2}.\label{AdSPoincare}
\ee
The insertion of the local operator $e^{-\epsilon H}\mathcal{O}(x_0)$ at $t=0$ corresponds to (is ``dual" to) a localised point particle in the bulk with mass $m\simeq \Delta$, for $\Delta\gg 1$ and $\epsilon\ll 1$, located at $x=x_0$ and $z=\epsilon$,  with vanishing velocity  \cite{Nozaki:2013wia,Berenstein:2019tcs}. As time evolves, this bulk particle follows a radial time-like geodesic $\{z(t),x=x_0\}$, as shown in figure \ref{fig:AdSP}.  \\
Geodesics are computed by extremising the world-line particle action
\begin{equation}
  S_\text{particle} = -m\int d\tau = -m\int d\lambda\,\sqrt{-g_{\mu \nu}(x)\dot{x}^\mu\dot{x}^\nu}\,,
\label{eq:p-action}
\end{equation}
describing the particle motion $x^\mu(\lambda)$ in a fixed spacetime metric with local components $g_{\mu \nu}(x)$. With the metric \eqref{AdSPoincare}, working in the gauge $\lambda=t$, \eqref{eq:p-action} becomes
\begin{equation}
    S_m=-m\int\frac{\sqrt{1-\dot{z}(t)^2}}{z(t)}dt.
 \label{SpPoincare}
\end{equation}
The particle following the on-shell geodesic satisfying our initial condition,
\be
z(t)=\sqrt{t^2+\epsilon^2}.\label{GeodesicPoincare}
\ee
carries the same energy as the CFT excitation \eqref{eq:energy}. 

An analogous picture holds for finite $\beta$, or $\beta=iL$, by replacing the bulk metric \eqref{AdSPoincare}, with an AdS black hole or global AdS, i.e. mapping the initial quantum state $|\psi_\beta\rangle$ to its proper semi-classical bulk description. In the black hole background, the point particle falls from the boundary towards the horizon region which is reached after the time scale $t\sim\beta/2\pi$ (see \eqref{eq:planar-BTZ}) \cite{Caputa:2014eta}.\\ 
\vspace{4 pt}
\noindent \textbf{\emph{4.~Spread Complexity of Local Operators.}} 
To compute the rate of growth of the spread complexity of states \eqref{LOES}, we first need to extract the Lanczos coefficients from the return amplitude. In our set-ups, return amplitudes can be universally determined in terms of ratios of two-point correlators of local operators $\mathcal{O}$ (see supplemental material A). At finite temperature, the amplitude is
\begin{equation}
S(t)=\left(\frac{\sinh\left(\frac{\pi(t+2i\epsilon)}{\beta}\right)}{\sinh\left(\frac{2\pi i\epsilon}{\beta}\right)}\right)^{-2\Delta}.\label{eq:RetAmpl}
\end{equation}
The corresponding Lanczos coefficients are given by
\be
a_n=\frac{2\pi(n+\Delta)}{\beta\tan\left(\frac{2\pi\epsilon}{\beta}\right)},\qquad b_n=\frac{\pi\sqrt{n(n+2\Delta-1)}}{\beta\sin\left(\frac{2\pi\epsilon}{\beta}\right)}.\label{LanczosBeta}
\ee
These are examples of the analytical Lanczos coefficients governed by $\SL(2,\mathbb{R})$ symmetry on the Krylov chain \cite{Caputa:2023vyr,Dymarsky:2019elm}. The corresponding wave functions can be computed (see \eqref{WFOps}), allowing us to find the spread complexity
\begin{equation}
    C_{\mt{K}}(t)=\frac{\Delta\beta^2}{2\pi^2\epsilon^2}\sinh^2\left(\frac{\pi t}{\beta}\right),
\end{equation}
and its rate of growth
\begin{equation}
    \dot{C}_{\mt{K}}(t)=\frac{E_0}{\epsilon}\frac{\beta}{2\pi}\sinh\left(\frac{2\pi t}{\beta}\right).\label{RatFinT}
\end{equation}
It is interesting to observe that the characteristic time scale $t\sim\beta/2\pi$ after which the rate grows exponentially matches the time it takes the dual point particle to reach the near-horizon region (see \eqref{eq:planar-BTZ}).\\
Proceeding in an analogous way, the growth rates of spread complexity for excited states \eqref{LOES} in a CFT of finite size $L$ and on an infinite line are
\be
\dot{C}_{\mt{K}}(t)=\frac{E_0}{\epsilon}\frac{L}{2\pi}\sin\left(\frac{2\pi t}{L}\right),\qquad \dot{C}_{\mt{K}}(t)=\frac{E_0}{\epsilon}t\,.
\label{ReatLInfL}
\ee
Together with \eqref{RatFinT}, these are our main CFT results for spread complexity rates. Next, we discuss how to match them with the proper momentum in AdS.\\
\vspace{4 pt}
\noindent \textbf{\emph{5.~Proper Bulk Momentum.}} 
Given a metric with radial coordinate $r$ (e.g. $z$ in \eqref{AdSPoincare}) and trajectory $r(t)$, the radial momentum is the canonical momentum
\begin{equation}
    P_r=\frac{\partial \mathcal{L}}{\partial \dot{r}(t)}.
\label{eq:rad-mom}
\end{equation}
Applying this definition to  \eqref{SpPoincare} and using \eqref{GeodesicPoincare}, one obtains
\be
P_z=\frac{m\,t}{\epsilon\sqrt{t^2+\epsilon^2}}.
\ee
Translating mass to CFT data, $m= \Delta$, we get a mismatch with the second formula in \eqref{ReatLInfL}. At early times the radial momentum grows linearly with $t$, but it approaches a constant at late times, unlike \eqref{ReatLInfL}.

Fortunately, there is an elegant way to correct this. The effective 2D metric probed by our infalling particle is locally AdS$_2$, whose isometry group $\SL(2,\mathbb{R})$ matches the symmetry controlling the Krylov states $\ket{K_n}$. Intuitively, since the spread complexity is basis-dependent, one may wonder whether a different radial coordinate could precisely encode the choice of the Krylov basis and the spread along the chain direction labeled by $n$. After all, small diffeomorphisms $z=f(r)$  preserve the $\SL(2,\mathbb{R})$ symmetry. Following this idea, covariance dictates $P_r = \frac{mt}{\epsilon}\frac{f^\prime(r)}{f(r)}$. Taking $z=e^{-\rho}$, we find that \eqref{eq:goal} indeed holds
\begin{equation}
  P_\rho=-\frac{m}{\epsilon}t \qquad \Rightarrow \qquad \dot{C}_K(t)=-\frac{P_\rho}{\epsilon}.
\label{eq:s1}
\end{equation}

We find an analogous radial coordinate $\rho$ for all the examples \eqref{LOES}. Namely, we can write the dual gravity backgrounds to the thermal state as well as to the finite and infinite size vacua as
\begin{align}
ds^2&=d\rho^2+\frac{4\pi^2}{\beta^2}\left(-\sinh^2(\rho)dt^2+\cosh^2(\rho)dx^2\right)\,,\label{ThermalProper}\\
ds^2&=d\rho^2+\frac{4\pi^2}{L^2}\left(-\cosh^2(\rho)dt^2+\sinh^2(\rho)d\phi^2\right)\,,\label{SizeLProper}\\
ds^2&=d\rho^2+e^{2\rho}\left(-dt^2+dx^2\right).\label{InfLineProper}
\end{align}
In all of them, the radial momentum \eqref{eq:rad-mom} conjugate to $\rho$ satisfies the second equation in \eqref{eq:s1}. Their common feature is that the radial coordinate $\rho$ computes the \emph{proper} distance to the black hole horizon \eqref{ThermalProper} (analogously to the center of global $AdS$ \eqref{ThermalProper} and to the Poincar\'e horizon \eqref{InfLineProper}, see supplemental material B). For this reason, we refer to $P_\rho$ as the \emph{proper momentum}.

To check the above claim, consider the time-like geodesic followed by an infalling particle in the black hole background \eqref{ThermalProper}
\be
\tanh\rho(t)=\frac{\tanh(\rho_\Lambda)}{\cosh\left(\frac{2\pi t}{\beta}\right)}.\label{GeodBH}
\ee
The particle's initial position is related to the $\epsilon$ regulator by $\rho_\Lambda=\log(\beta/(\pi\epsilon))$. It follows that the proper radial momentum is
\be
P_\rho=\frac{\partial\mathcal{L}}{\partial\dot{\rho}(t)}=-\frac{m}{\epsilon}\frac{\beta}{2\pi}\sinh\left(\frac{2\pi t}{\beta}\right).
\ee
An analogous computation can be performed for the massive particle in global $AdS_3$ with metric \eqref{SizeLProper}. The particle's trajectory in this case satisfies
\be
\tanh(\rho(t))=\cos\left(\frac{2\pi t}{L}\right)\tanh(\rho_\Lambda),\label{GeodGlob}
\ee
with $\rho_\Lambda=\log(L/(\pi\epsilon))$, leading to the proper momentum
\be
P_\rho=-\frac{m}{\epsilon}\frac{L}{2\pi}\sin\left(\frac{2\pi t}{L}\right).
\ee
In all three cases the momentum is negative, since the particle falls from the boundary at $\rho\sim\infty$ towards $\rho\sim 0$.\\
After mapping parameters to CFT data, we find a precise version of \eqref{eq:goal} relating the rate of growth of the spread complexity in 2D CFTs and the proper radial momentum of dual massive particles in the bulk
\begin{equation}
     \dot{C}_\mt{K}(t)=-\frac{1}{\epsilon}P_\rho(t).
\label{eq:final}
\end{equation}
The proportionality constant is the finite energy scale over which we regulate the operator. This is the main result of our work and is the first explicit check of the correspondence between the rate of growth of a well-defined complexity in 2D CFTs and the bulk radial momentum.\\
\vspace{4 pt}
\noindent \textbf{\emph{6.~Discussion and Outlook.}} 
We now discuss implications of our results and future directions.

First, the Euler-Lagrange equation for the infalling particle in \eqref{eq:goal} implies that the second derivative of the spread complexity
\begin{equation}
  \ddot{C}_{\mt{K}}(t) \propto \frac{d}{dt} \left(\frac{\partial {\cal L}}{\partial \dot{r}(t)}\right) = \frac{\partial {\cal L}}{\partial r},
\end{equation}
equals the force in a non-relativistic system (where ${\cal L} = T - V$ for a conservative force: $F(r) = -\frac{dV(r)}{dr}$). This ``Newton's law" relates the acceleration of the bulk particle with the acceleration of the spread complexity. This is consistent with the gravitational arguments in \cite{Susskind:2019ddc} and the Ehrenfest's theorem for spread complexity \cite{Erdmenger:2023wjg}.

Second, it is natural to speculate on the relation between the discrete Krylov label $n$ and the radial location $\rho$. As stressed in the main text, there is a single $\SL(2,\mathbb{R})$ symmetry determining the Krylov subspace and the 2D effective Lorentzian geometry probed by the infalling particle. This is remarkable since, due to factorisation, 2D CFTs are invariant under two copies of $\SL(2,\mathbb{R})$. The relation between these algebras and the Krylov one is in general non-trivial. This suggests
\begin{equation}
  |K_n\rangle = \sum_{r,s} \Gamma_{n,r,s} |h,r\rangle \otimes |\bar{h},s\rangle,
\end{equation}
with some non-trivial coefficients $\Gamma_{n,r,s}$.
Indeed, time evolution  in $\mathcal{H}_{\mt{CFT}}$ involves the spread of information in terms of left and right entangled excitations. Additionally, from the return amplitude perspective, one could interpret the excitation as a Gaussian around $x_0$ whose spread increases with time. That is the picture which is described in the bulk and is also captured by the symmetry of the Krylov subspace.

Third, moving away from our approximation, the infalling particle will back-react on the geometry. This will be captured by Einstein's equations
\be
R^{\mu\nu}-\frac{1}{2}Rg^{\mu\nu}+\Lambda g^{\mu\nu}=8\pi G_{\mt{N}} T^{\mu\nu},\label{EEBR}
\ee
where $T^{\mu\nu}$ is the particle's energy-momentum tensor computed from \eqref{eq:p-action}. In our set-up, we can check that the proper momentum is $P_\rho=\sqrt{-g}T^{0\rho}$. Hence, we can express our main result in terms of spacetime geometry data by either projecting \eqref{EEBR} on the above components or employing the gravitational constraints.  
This may bridge our work, which provides a sharp quantum mechanical definition of complexity in CFTs, with results relating the momentum of a radial shell of matter with the volume variation \cite{Barbon:2019tuq,Barbon:2020olv,Barbon:2020uux}. 

Fourth, our calculations relate exact quantum mechanical expectation values with classical conjugate momenta. The AdS/CFT correspondence and the symmetry controlling the dynamics of our examples provide some justification for the validity of the matching we reported. However, there is a broader perspective by which we can interpret our results. Since $C_{\mt{K}}(t)=\langle \psi(t)|\mathcal{K}|\psi(t)\rangle$ and $\mathcal{K} = L_0 - h\mathbf{1}$ within our set-up, it follows that
\begin{equation}
    \dot{C}_{\mt{K}}(t) \propto \bra{\psi(t)} [\mathcal{K}, H] \ket{\psi(t)}.
\end{equation}
Since $\mathcal{K},\,H\in \SL(2,\mathbb{R})\simeq \mathrm{SU}(1,1)$, it follows that the derivative $\dot{C}_{\mt{K}}(t) = \bra{\psi(t)} P \ket{\psi(t)}$ for some operator $P$ in this algebra. This result is exact and could be tested in other symmetry set-ups, including quantum optics, which makes heavy use of the formalism of $\mathrm{SU}(1,1)$ coherent states \cite{Agarwal_2012}.  

To test this identity and to partially link with the holographic logic, consider coherent states $\ket{\alpha}$ in a quantum harmonic oscillator. As shown in the supplemental material, the complexity operator is given by 
\begin{equation}\label{eq:coherent-krylov-operator-xp}
    \hat{\mathcal{K}} = \frac{( \hat{p} - \braket{p_0}_\alpha)^2}{2m\omega} + \frac{1}{2} m \omega (\hat{x} - \braket{x_0}_\alpha )^2\,,
\end{equation} 
and its first and second time derivatives, for coherent states satisfying the same initial condition as our infalling particles, $\braket{p_0}_\alpha=0$, are
\begin{equation}
  \left[\hat{\mathcal{K}}, \hat{H}\right] = -i\omega \braket{x_0}_\alpha\, \hat{p}\,, \quad \left[\left[\hat{\mathcal{K}},\hat{H}\right], \hat{H}\right] =-m\omega^3 \braket{x_0}_\alpha \hat{x}.
\label{eq:HO-t}
\end{equation}
These are operator equalities that hold beyond expectation values. However, since the harmonic oscillator potential is quadratic, Ehrenfest's theorem ensures that quantum expectation values satisfy the classical equations of motion. Thus, if we had approximated the exact quantum description by a classical one in the right hand side of these equations, we would have reached the same conclusion. We are envisioning a similar situation holds in our holographic discussion, modulo subtleties due to gravitational physics.

Finally, the relation between the spread complexity and proper radial momentum provides a guiding principle for holographic complexity and operator growth. The examples presented here are only starting points and we expect our relation to hold, at a minimum, in other quench scenarios and in higher dimensional CFTs. We will explore several of these directions in \cite{long-paper}.

\subsection*{Acknowledgements} 
We are grateful to Vijay Balasubramanian, Giuseppe Di Giulio, Javier Mag\'an, Dimitrios Patramanis and Onkar Parrikar for feedback and comments. JS was supported by the Science and Technology Facilities Council [grant number ST/X000494/1]. BS and RWM are supported by Engineering and Physical Sciences Research Council studentships. PC and BC are supported by NCN Sonata Bis 9 2019/34/E/ST2/00123 grant. PC is supported by the ERC Consolidator grant (number: 101125449/acronym: QComplexity). Views and opinions expressed are however those of the authors only and do not necessarily reflect those of the European Union or the European Research Council. Neither the European Union nor the granting authority can be held responsible for them.

\bibliography{ComplexityMomentum}

\pagebreak

\appendix

\titlepage

\setcounter{page}{1}

\begin{center}
{\LARGE Supplemental Material}
\end{center}

\section{A. Spread complexity and $\SL(2,\mathbb{R})$}\label{AppendixA}
The purpose of this appendix is to present the $\SL(2,\mathbb{R})$ Lie algebra methods used to find solutions to the Lanczos algorithm and to compute spread complexity in the main text. More detailed and pedagogical discussions can be found in \cite{Caputa:2021sib,Balasubramanian:2022tpr}.   

The generators $L_n$ of the $\SL(2,\mathbb{R})$ Lie algebra satisfy the commutation relations
\be
[L_n,L_m]=(n-m)L_{n+m},\qquad n,m=-1,0,1.
\ee
Since $L_{-1}$ and $L_{1}$ act like the raising and lowering operators, one defines the highest (lowest) weight state $\ket{h}$ as the eigenstate of $L_0$
\be
L_0\ket{h}=h\ket{h},\qquad L_1\ket{h}=0.
\ee
The natural orthonormal basis of states obtained by acting on $\ket{h}$ with raising operators is
\be
\ket{h,n}=\sqrt{\frac{\Gamma(2h)}{n!\Gamma(2h+n)}}L^n_{-1}\ket{h},\qquad \langle h,n|h,m\rangle=\delta_{n,m},\label{BasisSL2R}
\ee
whereas the algebra generators act on it as follows
\bea
L_0\ket{h,n}&=&(h+n)\ket{h,n},\nonumber\\
L_{1}\ket{h,n}&=&\sqrt{n(n+2h-1)}\ket{h,n-1},\nonumber\\
L_{-1}\ket{h,n}&=&\sqrt{(n+1)(n+2h)}\ket{h,n+1}.\label{ActionBasis}
\eea

Consider the real time evolution of the initial state $\ket{h}$ by a hypothetical Hamiltonian constructed from the Lie algebra generators as follows
\be
H=\gamma L_0+\alpha(L_1+L_{-1}).
\label{HamSL2}
\ee
Using the Baker-Campbell-Hausdorff (BCH) formula for $\SL(2,\mathbb{R})$ and expanding its action on the highest weight state in powers of the raising operator, one obtains
\bea
\ket{\psi(t)}\equiv e^{-iHt}\ket{h}=e^{hB}\sum^\infty_{n=0}A^nL^n_{-1}\ket{h}
=e^{hB}\sum^\infty_{n=0}\sqrt{\frac{\Gamma(2h+n)}{n!\Gamma(2h)}}A^n\ket{h,n},\label{StateSL2R}
\eea
where the basis \eqref{BasisSL2R} was used in the last equality.

Either by using general formulas or deriving them with an explicit representation, the BCH coefficients read
\be
A=-\frac{2i\alpha}{\mathcal{D}}\frac{\tanh\left(\frac{\mathcal{D}}{2}t\right)}{1+\frac{i\gamma}{\mathcal{D}}\tanh\left(\frac{\mathcal{D}}{2}t\right)},\qquad B=-2\log\left[\cosh\left(\frac{\mathcal{D}}{2}t\right)+\frac{i\gamma}{\mathcal{D}}\sinh\left(\frac{\mathcal{D}}{2}t\right)\right],
\ee
where a key role is played by the parameter
\be
\mathcal{D}=\sqrt{4\alpha^2-\gamma^2}.
\ee
If one interprets the Hamiltonian evolution as a vector flow, $\mathcal{D}$ is its norm. It can be real, imaginary or zero, depending on the values of $\gamma$ and $\alpha$.

Observe the Hamiltonian \eqref{HamSL2} is tri-diagonal when acting on the basis \eqref{BasisSL2R}. Indeed, using \eqref{ActionBasis}, one finds
\be
H\ket{h,n}=a_n\ket{h,n}+b_n\ket{h,n-1}+b_{n+1}\ket{h,n+1},\label{HinLieBasis}
\ee
with
\be
a_n=\gamma(h+n),\qquad b_n=\alpha\sqrt{n(n+2h-1)}.\label{LanczosSL@R}
\ee
These identities allow us to identify $\ket{h,n}$ as an example of the $n$th Krylov basis vector $\ket{K_n}$. Given the tri-diagonal action \eqref{HinLieBasis} of the Hamiltonian in that basis, which is the defining property for such basis in the Lanczos algorithm \cite{LanczosBook} , this provides an explicit example of Lanczos coefficients \eqref{LanczosSL@R}.  Moreover, from \eqref{StateSL2R}, one can read-off the wave functions $\psi_n(t)$ and probability distribution $|\psi_n(t)|^2$ within this $\SL(2,\mathbb{R})$ set-up:
\be
\psi_n(t)=\sqrt{\frac{\Gamma(2h+n)}{n!\Gamma(2h)}}e^{h B}A^n,\qquad \sum^\infty_{n=0}|\psi_n(t)|^2=1.\label{WFGen}
\ee
Together with \eqref{LanczosSL@R}, they solve the Schr\"odinger equation \eqref{SE1} from the main text.

The above discussion holds for arbitrary parameters $\gamma$ and $\alpha$ encoding the information about the initial state and the choice of Hamiltonian that is represented in our Krylov basis by \eqref{HamSL2}. Not every quantum evolution of some initial state with a general Hamiltonian can be represented by this emergent dynamical $\SL(2,\mathbb{R})$ symmetry. 

However, all the three CFT excited states \eqref{LOES} considered in this work can. This is the connection allowing us to perform analytical computations. Using the finite temperature $(\beta)$ case as an example, one has $h=\Delta$ and the Lanczos coefficients \eqref{LanczosBeta} have the $\SL(2,\mathbb{R})$ form \eqref{LanczosSL@R} with
\be
\gamma=\frac{2\pi}{\beta\tan\left(\frac{2\pi\epsilon}{\beta}\right)},\qquad \alpha=\frac{\pi}{\beta\sin\left(\frac{2\pi\epsilon}{\beta}\right)},\qquad \mathcal{D}=\frac{2\pi}{\beta}.
\ee
Note that the norm of the vector field coincides with the value of maximal Lyapunov exponent, as already observed and emphasized in \cite{Parker:2018yvk}. The wave functions \eqref{WFGen} simplify to 
\be
\psi_n(t)=\sqrt{\frac{\Gamma(2\Delta+n)}{n!\Gamma(2\Delta)}}\left(-\frac{\sinh\left(\frac{\pi(t-2i\epsilon)}{\beta}\right)}{\sinh\left(\frac{2\pi i\epsilon}{\beta}\right)}\right)^{-2\Delta}\left(-\frac{\sinh\left(\frac{\pi t}{\beta}\right)}{\sinh\left(\frac{\pi (t-2i\epsilon)}{\beta}\right)}\right)^n,\label{WFOps}
\ee
In particular, the return amplitude in \eqref{eq:RetAmpl} equals $S(t)=\psi^*_0(t)$. The key feature of this amplitude is that it is determined by the conformal invariance. Indeed, by construction of \eqref{LOES}, we can write it in terms of the ratio of two point correlators at finite temperature \cite{Caputa:2023vyr}
\be
S(t)=\frac{\langle\psi(t)|\psi(0)\rangle}{\langle\psi(0)|\psi(0)\rangle}=\frac{\langle \mathcal{O}(z_1,\bar{z}_1)\mathcal{O}(z_2(t),\bar{z}_2(t))\rangle}{\langle \mathcal{O}(z_1,\bar{z}_1)\mathcal{O}(z_2(0),\bar{z}_2(0))\rangle},
\ee
where the smearing energy $\epsilon$ and time $t$ are put into the insertion points
\be
z_1=x_0+i\epsilon,\quad \bar{z}_1=x_0-i\epsilon,\quad
z_2(t)=x_0-i(\epsilon+it),\quad \bar{z}_2(t)=x_0+i(\epsilon+it).
\ee
The final result is derived using the general two-point correlator at finite temperature (analogous result follows from finite and infinite size CFT two-point functions)
\be
\langle \mathcal{O}(z_1,\bar{z}_1)\mathcal{O}(z_2,\bar{z}_2\rangle=\left(\frac{\beta}{2\pi}\sinh\left(\frac{\pi z_{12}}{\beta}\right)\right)^{-2h}\left(\frac{\beta}{2\pi}\sinh\left(\frac{\pi \bar{z}_{12}}{\beta}\right)\right)^{-2\bar{h}},\qquad z_{ij}\equiv z_i-z_j.
\ee

Given the exact results reviewed in this appendix and the general remarks concerning the first and second time derivatives of the spread complexity written in the discussion and outlook section of the main text, we can confirm these by evaluating the expectation values of the $\SL(2,\mathbb{R})$ generators in our state \eqref{StateSL2R}
\bea
\langle\psi(t)|L_0|\psi(t)\rangle&=&\Delta\frac{1+|A|^2}{1-|A|^2}=2\Delta\frac{\sinh^2\left(\frac{\pi t}{\beta}\right)}{\sin^2\left(\frac{2\pi \epsilon}{\beta}\right)}+\Delta,\nonumber\\
\langle\psi(t)|L_1|\psi(t)\rangle&=&\Delta\frac{2A}{1-|A|^2}=-2\Delta\frac{\sinh\left(\frac{\pi t}{\beta}\right)\sinh\left(\frac{\pi (t+2i\epsilon)}{\beta}\right)}{\sin^2\left(\frac{2\pi \epsilon}{\beta}\right)},\nonumber\\
\langle\psi(t)|L_{-1}|\psi(t)\rangle&=&\Delta\frac{2\bar{A}}{1-|A|^2}=-2\Delta\frac{\sinh\left(\frac{\pi t}{\beta}\right)\sinh\left(\frac{\pi (t-2i\epsilon)}{\beta}\right)}{\sin^2\left(\frac{2\pi \epsilon}{\beta}\right)}.\label{ExpectationValues}
\eea
Analogous computations can be carried out for CFTs at finite size $L$ (this is equivalent to setting $\beta=iL$), as well as for CFTs on an infinite line (extracted from \eqref{ExpectationValues} by $\beta\to\infty$). In all these computations we can verify that the actual spread complexity can be written as \cite{Caputa:2021ori}
\be
C_{\mt{K}}(t)=\langle\psi(t)|L_0|\psi(t)\rangle-\langle\psi(0)|L_0|\psi(0)\rangle.
\ee
It follows that its first derivative can be written as the expectation value of the operator $\alpha(L_1-L_{-1}) \in \SL(2,\mathbb{R})$ 
\be
\dot{C}_{\mt{K}}(t)=i\langle\psi(t)|[H,L_0]|\psi(t)\rangle=i\alpha\langle\psi(t)|(L_1-L_{-1})|\psi(t)\rangle,
\ee
while its second derivative is also fixed by the underlying  $\SL(2,\mathbb{R})$ algebra 
\be
\ddot{C}_{\mt{K}}(t)=i^2\langle\psi(t)|[H,[H,L_0]]|\psi(t)\rangle=4\alpha^2\langle\psi(t)|L_0|\psi(t)\rangle+\gamma\alpha\langle\psi(t)|L_1+L_{-1}|\psi(t)\rangle.
\ee
This indeed matches with the second derivative in our examples
\be
\ddot{C}_{\mt{K}}(t)=
\frac{E_0}{\epsilon}\cosh\left(\frac{2\pi t}{\beta}\right).
\ee
These derivatives are examples of the Ehrenfest's theorem for Krylov/spread complexity discussed in \cite{Erdmenger:2023wjg}.

\section{B. Embeddings and proper distance.}
\label{AppendixB}
The purpose of this appendix is to introduce the different AdS$_3$ coordinate systems used in the main text together with the definition of the proper distance.\\

3D Anti-de Sitter is a maximally symmetric space defined as the universal cover of the quadratic surface
\begin{equation}
  \eta_{\mt{AB}}X^\mt{A} X^\mt{B} = -(X^0)^2 - (X^1)^2 + (X^2)^2 + (X^3)^2 = -\ell^2,
\label{eq:quadratic}
\end{equation}
embedded in $\mathbb{R}^{2,2}$. The geodesic distance $D(X,Y)$ between two time-like separated points $X^\mt{A}$ and $Y^\mt{A}$ is given by
\begin{equation}
  \cosh D(X,Y) = -\frac{\eta_{\mt{AB}} X^\mt{A}Y^\mt{B}}{\ell^2}\,.
\label{eq:g-distance}
\end{equation}
Below, we set the radius $\ell$ of AdS$_3$ to one.

\subsection{Black Hole}
The semi-classical gravity description of the 2D CFT at finite temperature on the real line involves the planar version of the three-dimensional non-rotating black hole spacetime  \cite{Banados:1992wn}. The region outside its horizon can be described by two embeddings
\begin{equation}
\begin{aligned}
    X^0=\frac{\sqrt{1-Mz^2}}{\sqrt{M}z}\sinh(\sqrt{M}t)=\sinh(\rho)\sinh\left(\frac{2\pi}{\beta}t\right), \quad &
    \quad X^1=\frac{1}{\sqrt{M}z}\cosh(\sqrt{M}x)=\cosh(\rho)\cosh\left(\frac{2\pi}{\beta}x\right), \\
    X^2=\frac{1}{\sqrt{M}z}\sinh(\sqrt{M}x)=\cosh(\rho)\sinh\left(\frac{2\pi}{\beta}x\right),\quad & \quad
    X^3=\frac{\sqrt{1-Mz^2}}{\sqrt{M}z}\cosh(\sqrt{M}t)=\sinh(\rho)\cosh\left(\frac{2\pi}{\beta}t\right)\,.
\end{aligned}
\label{EmbeddingBH}
\end{equation}
The first embedding corresponds to the planar BTZ metric
\be
ds^2=\frac{1}{z^2}\left(-(1-Mz^2)dt^2+\frac{dz^2}{1-Mz^2}+dx^2\right),
\ee
with inverse temperature $\beta=T^{-1}=2\pi/\sqrt{M}$ determining its horizon at $z_\star=\beta/2\pi$.\\ 
In this geometry, the massive point particle describing the dual to the local 2D CFT quench follows the trajectory \cite{Caputa:2014eta}
\be
z(t)=\frac{\beta}{2\pi}\sqrt{1-\left(1-\frac{4\pi^2 \epsilon^2}{\beta^2}\right)\left(1-\tanh^2\left(\frac{2\pi t}{\beta}\right)\right)}.
\label{eq:planar-BTZ}
\ee
It starts at the cut-off surface $z(0)=\epsilon$ and falls towards the horizon. After the time scale $t\simeq \beta/2\pi$, one can approximate $\tanh\left(\frac{2\pi t}{\beta}\right)\sim 1$, and the particle explores the near horizon region of the black hole.\\
The second embedding in \eqref{EmbeddingBH} yields the metric
\be
ds^2=d\rho^2+\frac{4\pi^2}{\beta^2}\left(-\sinh^2(\rho)dt^2+\cosh^2(\rho)dx^2\right)\,,
\ee
matching \eqref{ThermalProper} in the main text. The black hole horizon is located at $\rho=0$, while the AdS$_3$ conformal boundary is at $\rho\to\infty$. The bulk particle's trajectory is given by \eqref{GeodBH}. The embedding formulas \eqref{EmbeddingBH} relate the initial location of the particle $\rho_\Lambda$ with the regulator $\epsilon$
\be
e^{\rho_\Lambda}=\frac{\beta}{\pi\epsilon}.
\ee
Finally, the geodesic distance \eqref{eq:g-distance} between two points in these coordinates reads
\be
\cosh (D_{12})=\cosh(\rho_1)\cosh(\rho_2)\cosh\left(\frac{2\pi\Delta x}{\beta}\right)-\sinh(\rho_1)\sinh(\rho_2)\cosh\left(\frac{2\pi\Delta t}{\beta}\right),
\ee
where $\Delta t\equiv t_1-t_2$ (and similarly for the other coordinates). Thus, for points on the same time and space slice ($\Delta t=\Delta x=0$), it follows
\be
\cosh (D_{12})=\cosh(\Delta\rho),
\ee
the coordinate $\rho$ matches the proper distance to the horizon at $\rho=0$.
\subsection{Global coordinates}
The semi-classical gravity dual to the 2D CFT vacuum on a circle of length $L$ is given by global AdS$_3$. This corresponds to the embeddings 
\begin{equation}
    \begin{aligned}
        X^0=\sqrt{1+r^2}\cos(t)=\cosh(\rho)\cos\left(\frac{2\pi t}{L}\right),\quad & \quad
        X^1=\sqrt{1+r^2}\sin(t)=\cosh(\rho)\sin\left(\frac{2\pi t}{L}\right), \\
        X^2=r\sin(\phi)=\sinh(\rho)\sin\left(\frac{2\pi \phi}{L}\right),\quad & \quad
        X^3=r\cos(\phi)=\sinh(\rho)\cos\left(\frac{2\pi \phi}{L}\right).
    \end{aligned}
\label{EmbeddingGlobal}
\end{equation}
The first embedding yields the global AdS$_3$ metric 
\be
ds^2=-f(r)dt^2+\frac{dr^2}{f(r)}+r^2D\phi^2,\qquad f(r)=1+r^2,
\ee
whereas the second reproduces \eqref{SizeLProper}
\be
ds^2=d\rho^2+\frac{4\pi^2}{L^2}\left(-\cosh^2(\rho)dt^2+\sinh^2(\rho)d\phi^2\right).
\ee

In the first geometry, the massive point particle describing the dual to the local 2D CFT quench follows the trajectory
\be
r(\tau)=\frac{r_\Lambda\cos(\tau)}{\sqrt{r_\Lambda\sin^2(\tau)+1}},
\ee
while the trajectory is given by \eqref{GeodGlob} for the second metric. Comparing the asymptotic growth of the metric close to the boundary determines the relation between its initial position $\rho_\Lambda$ and the regulator $\epsilon$ to be
\begin{equation}
    e^{\rho_\Lambda}=\frac{L}{\pi\epsilon}
\end{equation}
The interpretation for the $\rho$ coordinate can be inferred from the geodesic distance \eqref{eq:g-distance} between two points
\be
\cosh (D)=\cosh(\rho_1)\cosh(\rho_2)\cos\left(\frac{2\pi\Delta t}{L}\right)-\sinh(\rho_1)\sinh(\rho_2)\cos\left(\frac{2\pi\Delta \phi}{L}\right),
\ee
so that for $\Delta t=\Delta\phi=0$ coordinate $\rho$ is the proper distance in the bulk to the origin of the AdS$_3$ spacetime.
\subsection{Poincar\'e patch}
The semi-classical gravity dual to the 2D CFT vacuum on the infinite line is given by the Poincar\'e patch of AdS$_3$. This corresponds to the embedding
\begin{equation}
\begin{aligned}
    X^0=\frac{1-t^2+x^2+z^2}{2z}=\frac{1}{2}e^\rho\left(1-t^2+x^2+e^{-2\rho}\right), \quad & \quad
    X^1=\frac{t}{z}=e^\rho t,\\
    X^3=\frac{-1-t^2+x^2+z^2}{2z}=\frac{1}{2}e^\rho\left(-1-t^2+x^2+e^{-2\rho}\right) \quad & \quad 
    X^2=\frac{x}{z}=e^\rho x.
\end{aligned}
\label{EmbeddingPoincare}     
\end{equation}
where we used $z=e^{-\rho}$ in the second equality. The first choice reproduces the Poincar\'e metric \eqref{AdSPoincare}
\be
ds^2=\frac{-dt^2+dx^2+dz^2}{z^2},
\ee
whereas the second gives rise to \eqref{InfLineProper}
\be
ds^2=d\rho^2+e^{2\rho}(-dt^2+dx^2).
\ee

The massive point particle in this geometry describing the dual to the local 2D CFT quench follows the trajectory
\eqref{GeodesicPoincare}. Written in terms of $\rho(t)$,
\be
\rho(t)=-\frac{1}{2}\log\left(t^2+\epsilon^2\right).
\ee
The interpretation for the $\rho$ coordinate can be inferred from the geodesic distance \eqref{eq:g-distance} between two points
\be
\cosh (D_{12})=\cosh(\Delta\rho)+\frac{1}{2}e^{\rho_1+\rho_2}\left(-\Delta t^2+\Delta x^2\right).
\ee
For $\Delta t=\Delta x=0$, the geodesic distance is controlled by variations in $\rho$, confirming its proper radial distance interpretation.

\section{C. Coherent states in a harmonic oscillator }
\label{AppendixC}
The purpose of this appendix is to compute the complexity operator \eqref{eq:coherent-krylov-operator-xp} for coherent states of the quantum harmonic oscillator. The Hamiltonian operator $\hat{H} = \frac{\hat{p}^2}{2m} + \frac{1}{2}m\omega^2\hat{x}^2$ can be equivalently written in terms of the number operator $\hat{N} = \hat{a}^\dagger \hat{a}$ as $\hat{H}=\omega\left(\hat{N} + \frac{1}{2}\right)$. Together with the ladder operators
\begin{equation}
  \hat{a} = \sqrt{\frac{m\omega}{2}} \left(\hat{x} + \frac{i}{m\omega}\,\hat{p}\right)\,, \quad \quad \hat{a}^\dagger = \sqrt{\frac{m\omega}{2}} \left(\hat{x} - \frac{i}{m\omega}\,\hat{p}\right)\,,
\label{eq:ladder}
\end{equation}
these satisfy the Heisenberg-Weyl algebra
\begin{equation}
  \left[\hat{a},\,\hat{a}^\dagger\right] = \mathbb{I}\,, \quad \left[\hat{N},\,\hat{a}^\dagger\right] = \hat{a}^\dagger\,, \quad \quad \left[\hat{N},\,\hat{a}\right] = -\hat{a}\,.
\label{eq:HW-algebra}
\end{equation}
This algebra has a representation on the familiar number basis $\ket{n}$, with
\begin{equation}
  \hat{N}\ket{n} = n\ket{n}\,, \qquad \hat{a}^\dagger \ket{n} = \sqrt{n+1}\ket{n+1}\,, \qquad \hat{a}\ket{n} = \sqrt{n}\ket{n-1}\,.
\end{equation}

Coherent states $\ket{\alpha}$ are eigenstates of the ladder operator $\hat{a}\ket{\alpha} = \alpha\ket{\alpha}$, with $\alpha \in \mathbb{C}$. Their expansion in the number basis
\begin{equation}
  \ket{\alpha}= \exp\left(-\frac{|\alpha|^2}{2}\right) \sum_{n=0}^\infty \frac{\alpha^n}{\sqrt{n!}} \ket{n},
\end{equation}
allows us to compute their time evolution
\begin{equation}\label{eq:coherent-state-time}
    e^{-i \hat{H} t} \ket{\alpha} = e^{-i \omega t/2} \ket{\alpha e^{-i \omega t}}\,.
\end{equation}
It follows that the return amplitude equals
\begin{equation}
  S(t) = \braket{\alpha|e^{i \hat{H} t}|\alpha} = e^{i \omega t/2} e^{-|\alpha|^2} \sum_n \frac{(\alpha^* \alpha e^{i \omega t})^n}{n!} = e^{i \omega t/2 + |\alpha|^2 (e^{i \omega t}-1)}.
\end{equation}
Following the algorithm described in the main text, one can compute the Lanczos coefficients from the moments of $S(t)$, solve for the wave functions $\psi_n(t)$ and derive the spread complexity, as presented in \cite{Balasubramanian:2022tpr}. Here, we present an alternative derivation.

A coherent state $\ket{\alpha}$ can be prepared by acting with the \emph{displacement operator} $\hat{D}(\alpha)$ on the oscillator vacuum $|0\rangle$
\begin{equation}\label{eq:displacement-operator}
    \ket{\alpha} = \hat{D}(\alpha) \ket{0},
    \qquad \hat{D}(\alpha) = \exp(\alpha \hat{a}^\dagger - \alpha^* \hat{a}), \qquad \alpha \in\mathbb{C}.
\end{equation}
$\hat{D}(\alpha)$ is unitary, $\hat{D}(\alpha)\hat{D}^\dagger(\alpha) = \hat{D}^\dagger(\alpha) \hat{D}(\alpha)=\mathbb{I}$, and it acts on the ladder operators as
\begin{equation}\label{eq:displacement-ladder}
    \hat{D}(\alpha)^\dagger \, \hat{a} \, \hat{D}(\alpha) = \hat{a} + \alpha\,\mathbb{I}, \qquad
    \hat{D}(\alpha)^\dagger \, \hat{a}^\dagger \, \hat{D}(\alpha) = \hat{a}^\dagger + \alpha^*\,\mathbb{I}.
\end{equation}
The action of $\hat{H}$ on $\ket{\alpha}$ is unitarily equivalent to the action of a modified Hamiltonian $\tilde{H}$ on the usual number basis
\begin{equation}
  \hat{H} \ket{\alpha} = \hat{D}(\alpha) \hat{D}(\alpha)^\dagger \hat{H} \hat{D}(\alpha) \ket{0} = \hat{D}(\alpha) \tilde{H} \ket{0}.
\end{equation}
Using \eqref{eq:displacement-ladder}, $\tilde{H}$ equals
\begin{equation}\label{eq:equivalent-hamiltonian}
    \tilde{H} = \hat{D}(\alpha)^\dagger \hat{H} \hat{D}(\alpha) = \omega((\hat{a}^\dagger + \alpha^*\,\mathbb{I})(\hat{a}+\alpha\,\mathbb{I})+\frac{1}{2}) = \omega\left(\hat{N} + |\alpha|^2 + \frac{1}{2} + \alpha^* \hat{a} + \alpha \hat{a}^\dagger\right).
\end{equation}
It follows the action of $\tilde{H}$ on the number basis equals
\begin{equation}
  \tilde{H} \ket{n} = \omega(|\alpha|^2+\frac{1}{2}+n) \ket{n}  + \omega \alpha^* \sqrt{n} \ket{n-1} + \omega \alpha \sqrt{n+1} \ket{n+1}\,.
\end{equation}
Comparing the latter with the tri-diagonal action of the Hamiltonian in the Krylov basis \eqref{HinLieBasis} identifies the Krylov basis vectors $\ket{\tilde{K}_n}$ and the Lanczos coefficients to be
\begin{equation}
\label{eq:action-on-n-basis}
    \ket{\tilde{K}_n} = e^{i n \arg \alpha} \ket n, \quad a_n = \omega(|\alpha|^2+\frac{1}{2}+n), 
    \quad b_n = \omega|\alpha| \sqrt{n}.
\end{equation}
Notice, in particular, how the phase defining the Krylov basis is required by $\alpha \in \mathbb{C}$. However, when evaluating the Krylov operator $\tilde{\mathcal{K}}$, the dependence on this phase cancels and $\tilde{\mathcal{K}}$ equals the number operator
\begin{equation}\label{eq:krylov-operator-n-basis}
    \tilde{\mathcal{K}} = \sum_n n \ket{\tilde{K}_n} \bra{\tilde{K}_n} = \sum_n n \ket{n} \bra{n} = N.
\end{equation}

Our goal is to compute the complexity operator $\hat{\mathcal{K}}$ associated with the original Hamiltonian $\hat{H}$. This can be computed by returning to the coherent state basis by applying the displacement operator. It follows
\begin{equation}\label{eq:krylov-operator-coherent-basis}
    \hat{\mathcal{K}} = \hat{D}(\alpha) \hat{a}^\dagger \hat{D}^\dagger(\alpha)\,\hat{D}(\alpha) \hat{a} \hat{D}^\dagger(\alpha) = (\hat{a}^\dagger-\alpha^*\,\mathbb{I})(\hat{a}-\alpha\,\mathbb{I}),
\end{equation}
and the Krylov basis vectors are given by
\begin{equation}\label{eq:krylov-basis-coherent}
    \ket{K_n} = \hat{D}(\alpha) \ket{\tilde{K}_n} = e^{i n \arg \alpha} \hat{D}(\alpha) \ket n.
\end{equation}
Using \eqref{eq:ladder}, one can write \eqref{eq:krylov-operator-coherent-basis} in terms of the standard position and momentum operators as
\begin{equation}
    \hat{\mathcal{K}} = \frac{(\hat{p}-\braket{p_0}_\alpha )^2}{2m \omega} + \frac{1}{2} m \omega (\hat{x}-\braket{x_0}_\alpha)^2,
\end{equation}
where $\braket{x_0}_\alpha, \braket{p_0}_\alpha$ are the expectation values of the position and momentum in the initial coherent state $\ket{\alpha}$. This reproduces  statement \eqref{eq:coherent-krylov-operator-xp} in the main body of this letter.

The time derivatives of the complexity operator are fully determined by the Heisenberg-Weyl algebra \eqref{eq:HW-algebra}. Indeed,
\begin{equation}
  \left[\hat{\mathcal{K}}, \hat{H}\right] = \omega\left(\alpha \hat{a}^\dagger - \alpha^\star \hat{a}\right)\,, \qquad
  \left[\left[\hat{\mathcal{K}},\hat{H}\right], \hat{H}\right] = -\omega^2\left(\alpha\,\hat{a}^\dagger + \alpha^\star\,\hat{a}\right)\,.
\end{equation}
In particular,  when $\alpha=\alpha^\star$, the initial momentum expectation value vanishes $(\braket{p_0}_\alpha = 0)$, and these derivatives reduce to
\begin{equation}
    \left[\hat{\mathcal{K}}, \hat{H}\right] = -i\omega \braket{x_0}_\alpha\, \hat{p}\,, \qquad \left[\left[\hat{\mathcal{K}},\hat{H}\right], \hat{H}\right] = -m\omega^3 \braket{x_0}_\alpha \hat{x},
\end{equation}
reproducing the main text statements \eqref{eq:HO-t}.

\end{document}